\documentclass[prd,twocolumn,aps,showpacs,nofootinbib,nobibnotes,superscriptaddress,preprintnumbers]{revtex4}
\usepackage{epsfig}
\usepackage{graphics}
\usepackage{bm}
\usepackage{color}
\usepackage{dcolumn}   
\usepackage{bm}     
\usepackage{bbm}       
\usepackage{amssymb}  
\usepackage{amsmath}
\usepackage{latexsym}
\usepackage{float}
\usepackage{ifthen}
\usepackage{caption,subfig}
\usepackage{enumerate}
\usepackage{url}
\usepackage{caption,subfig}
\usepackage{amsopn}
\usepackage{hyperref}

\usepackage{amsfonts}
\usepackage{multirow}
\usepackage{array}
\usepackage{booktabs}
\usepackage{rotating}

\def\clap#1{\hbox to 0pt{\hss#1\hss}}

\def\({\left(}
\def\){\right)}
\def\[{\left[}
\def\]{\right]}
\def\bea{\begin{eqnarray}}
\def\eea{\end{eqnarray}}
\def\be{\begin{equation}}
\def\ee{\end{equation}}
\def\ba{\begin{eqnarray}}
\def\ea{\end{eqnarray}}
\def\beq{\begin{eqnarray}}
\def\eeq{\end{eqnarray}}
\def\mpl{M_{\rm Pl}}
\def\d{\mathrm{d}}
\newcommand{\cs}{c_s}
\newcommand{\lambdat}{\tilde{\lambda}}

\newcommand{\Mm}{\hat{M}}
\newcommand{\Nm}{\hat{N}}
\newcommand{\Wm}{\hat{W}}
\newcommand{\Fm}{\hat{F}}
\newcommand{\Rm}{\hat{R}}
\newcommand{\gm}{\hat{g}}

\newcommand{\Tm}{\hat{T}}

\newcommand{\gt}{\tilde{g}}
\newcommand{\hT}{\tilde{h}}
\newcommand{\hTc}{\tilde{\texttt{h}}}
\newcommand{\Ht}{\tilde{H}}

\newcommand{\at}{\tilde{a}}
\newcommand{\nt}{\tilde{n}}
\newcommand{\Id}{\mathbbm 1}

\newcommand{\mS}{\mathcal{S}}
\newcommand{\Od}{\mathcal{O}}

\def\cs{c_{\rm s}}

\def\be{\begin{equation}}
\def\ee{\end{equation}}
\def\ba{\begin{eqnarray}}
\def\ea{\end{eqnarray}}
\def\beq{\begin{eqnarray}}
\def\eeq{\end{eqnarray}}

\def\mpl{M_{\rm Pl}}

\def\d{\mathrm{d}}

\def\L*{{\cal L}_*}
\def\L{\mathcal{L}}
\def\({\left(}
\def\){\right)}

\def\<{\langle}
\def\>{\rangle}

\def\cs2{c_{s}^{2}}

\def\be{\begin{equation}}
\def\ee{\end{equation}}
\def\ba{\begin{eqnarray}}
\def\ea{\end{eqnarray}}
\def\beq{\begin{eqnarray}}
\def\eeq{\end{eqnarray}}

\def\mpl{M_{\rm Pl}}

\def\d{\mathrm{d}}

\def\L*{{\cal L}_*}
\def\L{\mathcal{L}}
\def\({\left(}
\def\){\right)}

\def\<{\langle}
\def\>{\rangle}

\begin{document}
\hspace{5.2in} \mbox{NORDITA-2015-36}\\\vspace{1.53cm} 

\title{Tensor perturbations in a general class of Palatini theories}

\date{\today,~ $ $}

\author{Jose Beltr\'an Jim\'enez}
\email{jose.beltran@cpt.univ-mrs.fr}
\affiliation{Centre for Cosmology, Particle Physics and Phenomenology,
Institute of Mathematics and Physics, Louvain University,
2 Chemin du Cyclotron, 1348 Louvain-la-Neuve, Belgium.}
\affiliation{CPT, Aix Marseille Universit\'e, UMR 7332, 13288 Marseille,  France.}

\author{Lavinia Heisenberg}
\email{laviniah@kth.se}
\affiliation{Nordita, KTH Royal Institute of Technology and Stockholm University, Roslagstullsbacken 23, 10691 Stockholm, Sweden}
\affiliation{Department of Physics \& The Oskar Klein Centre, AlbaNova University Centre, 10691 Stockholm, Sweden}

\author{Gonzalo J.  Olmo}
\email{gonzalo.olmo@csic.es}
\affiliation{Depto. de F\'{i}sica Te\'{o}rica \& IFIC, Universidad de Valencia - CSIC,Calle Dr. Moliner 50, Burjassot 46100, Valencia, Spain}
\affiliation{Depto. de F\'isica, Universidade Federal da Para\'\i ba,  Cidade Universit\'aria, s/n - Castelo Branco, 58051-900 Jo\~ao Pessoa, Para\'\i ba, Brazil}

\date{\today}

\begin{abstract}
We study a general class of gravitational theories formulated in the Palatini approach and derive the equations governing the evolution of tensor perturbations. In the absence of torsion, the connection can be solved as the Christoffel symbols of an auxiliary metric which is non-trivially related to the space-time metric. We then consider background solutions corresponding to a perfect fluid and show that the tensor perturbations equations (including anisotropic stresses) for the auxiliary metric around such a background take an Einstein-like form. This facilitates the study in a homogeneous and isotropic cosmological scenario where we explicitly establish the relation between the auxiliary metric and the space-time metric tensor perturbations. As a general result, we show that both tensor perturbations coincide in the absence of anisotropic stresses.
\end{abstract}
\maketitle

\section{Introduction}

Almost a century after its proposal, General Relativity (GR) is still considered as the standard theory for gravitational interactions. It is used to describe gravity in a wide range of scales where it has shown excellent agreement with experiments \cite{Will:2014xja}. This success has motivated its use in a broader range of scales. However, this extrapolation makes necessary to introduce not so well understood new physics to explain observations. Thus, when going to larger scales we encounter the necessity of adding a dark matter component in order to explain the rotation curves of galaxies, lensing measurements, large scale structure formation, etc. An even more intriguing component dubbed dark energy is required to explain the current phase of accelerated expansion. Although this cosmic acceleration can be easily justified by a simple cosmological constant, its unnatural observed value and its instability against typical quantum corrections also makes it necessary to go beyond GR and/or standard quantum field theory. Furthermore, the presence of cosmological and black hole singularities are typically seen as signals of the breakdown of GR. These singularities are usually expected to be regularized by a theory of quantum gravity.

Modifications of GR can be broadly divided into two approaches, namely: an effective field theory point of view and a more geometrical perspective. In the former, GR arises as the consistent theory for the interaction of matter through a massless spin-2 field. Thus, modifications of the gravitational interaction imply the introduction of additional degrees of freedom. These modifications lead to scalar-tensor theories \cite{scalartensor}, vector-tensor theories \cite{vectortensor}, massive gravity \cite{massivegravity}, etc. In the geometrical view, GR is a theory that describes the dynamics of the space-time geometry where one fixes the affine structure so that the connection is given by the Levi-Civita connection of the space-time metric, whose dynamics is then determined by Einstein's equations. Thus, if one intends to modify this theory, one can modify the field equations for the metric tensor or the gravitational interaction, like in the $f(R)$ theories \cite{f(R)}, braneworld models \cite{braneworld}, universal non-minimal couplings to matter \cite{nonminimalcouplings}. Another approach, however, consists in giving up on the Riemannian  geometry and consider more general geometric scenarios. In some cases, only a few degrees of freedom are allowed to propagate, like models with torsion \cite{torsion}, theories formulated in conformal or Weyl geometries \cite{Weyl}, bimetric variational principles for GR \cite{bimetric}, etc. In the most general case, the connection is considered as an independent object with no restrictions to be dynamically determined by the field equations, i.e., the metric and the connection are treated on equal footing. Theories within this framework are usually referred to as formulated \`a la Palatini \cite{Olmo2011}. Interestingly, a large class of theories formulated within this framework do not actually introduce additional propagating degrees of freedom since the connection can be regarded as an auxiliary field. Once the connection is integrated out, one is effectively left with GR  and the modifications are transferred to the couplings to matter.

In this work we will focus on theories in which the Einstein-Hilbert action is replaced by a function which depends on the possible contractions between the metric $g_{\mu\nu}$ and the Ricci tensor $R_{\mu\nu}(\Gamma)$ where the connection $\Gamma$ is independent of $g_{\mu\nu}$. Within this general class of modifications of GR, we work out a general approach for the computation of tensor perturbations and show the emergent parallels to GR. Theories of the form $f(R,R_{\mu\nu}R^{\mu\nu})$ and Born-Infeld inspired gravity theories are included in the general class of functions that we consider here and, thus, our results will be valid for them.

\section{Generalities}
Throughout this work we are going to consider the general class of gravitational theories in the Palatini formalism which are described by the following action
\be
\mS=\lambdat^4\int\d^4x\sqrt{-g}F(R_{\alpha\beta}(\Gamma), g^{\mu\nu})
\ee
where $F$ is an arbitrary function (that we will assume to be analytic), $R_{\alpha\beta}(\Gamma)$ is the Ricci tensor associated to an arbitrary connection $\Gamma^\alpha_{\mu\nu}$, $g^{\mu\nu}$ represents the inverse of the metric tensor and $\lambdat$ is some energy scale. Notice that we explicitly assume that the function $F$ only depends on the inverse of the metric and not on the metric itself. Since the action must be a scalar and the Ricci tensor needs not to be symmetric in general, it will be convenient to introduce the matrices $M^\mu{}_\nu\equiv \lambda^{-2}g^{\mu\alpha}R_{\alpha\nu}$ and $N^\mu{}_\nu\equiv \lambda^{-2}g^{\mu\alpha}R_{\nu\alpha}$, with $\lambda$ some energy scale\footnote{This energy scale is introduced for later convenience. In most cases it is determined by the specific functional form of the function $F$ and does not need to coincide with $\lambdat$.}, and rewrite the action in the form
\be\label{eq:action0}
\mS=\lambdat^4\int\d^4x\sqrt{-g}F(\Mm, \Nm) \ .
\ee
Given the scalar character of $F$, it can only depend on the traces of arbitrary products\footnote{If the action only depends on one of the matrices $\Mm$ or $\Nm$, then $F$ can only depend on traces of powers of the matrix. In such a case, we can go one step further and make use of the Cayley-Hamilton theorem for the corresponding matrix, say $\Mm$, which allows to express the traces of $[\Mm^n]$ with $n>4$ in terms of the traces $[\Mm]$, $[\Mm^2]$, $[\Mm^3]$ and $[\Mm^4]$ (and analogously for theories depending only on $\Nm$). As a result, we will have $F=F(X_1,X_2,X_3,X_4)$ where $X_n$ stands for the trace of the $n$-th power of $\Mm$ or $\Nm$. For this class of theories many of the expressions become greatly simplified.} of $\Mm$ and $\Nm$ and, in addition, it must satisfy the relation $F(\hat{A}^{-1}\Mm\hat{A},\hat{A}^{-1}\Nm\hat{A})=F(\Mm, \Nm)$ for any non-degenerate matrix $\hat{A}$. It will be useful for later purposes to note that the matrices $\Mm$ and $\Nm$ satisfy the relation $\Mm=\gm^{-1}\Nm^T\gm$. Only in the case when the Ricci tensor is symmetric, do $\Mm$ and $\Nm$ coincide. This simplified case is actually dynamically found in many physically relevant situations, as we shall show below. By distinguishing between $\Mm$ and $\Nm$, however, we put forward that a more general scenario is possible. \\

At low curvatures, the action (\ref{eq:action0}) can be Taylor-expanded as
\ba
\mS&=&\lambdat^4\int\d^4x\sqrt{-g}\left[F_0+\frac{\partial F}{\partial M^{\mu}{}_{\nu}}\Big\vert_0M^{\mu}{}_{\nu}+\frac{\partial F}{\partial N^{\mu}{}_{\nu}}\Big\vert_0N^{\mu}{}_{\nu}\right] \nonumber\\
&+&\Od(\lambda^{-2}\Rm^2) \ ,
\ea
where the subscript $0$ means evaluation at $\Mm=\Nm=0$. In order to recover GR in this limit, we need to impose
\be
\frac{\partial F}{\partial M^{\mu}{}_{\nu}}\Big\vert_0=\alpha\delta_\mu{}^\nu\quad{\rm and}\quad\frac{\partial F}{\partial N^{\mu}{}_{\nu}}\Big\vert_0=\beta\delta_\mu{}^\nu
\ee
with $2\lambdat^4\lambda^{-2}(\alpha+\beta)=\mpl^2$. In such a case, one recovers the  action of the Palatini version of GR  (with a possible cosmological constant), which is dynamically equivalent to the metric formulation. It should be noticed that the usual projective invariance of the Einstein-Hilbert action arises here only as an accidental symmetry, as higher order terms included in $\Od(\lambda^{-2}\Rm^2)$ will generically break it. Particular cases are those theories in which only the symmetric part of $R_{\mu\nu}$ appears in the action\footnote{The condition of having a symmetric Ricci tensor is not the same as having a torsion-free connection since the non-metricity tensor can also generate a non-symmetric part of the Ricci tensor.}, since, in those cases, the projective invariance is a symmetry of the full action and not only an accidental symmetry of the low curvature limit.

\section{Field equations}
Now that the action for a rather general class of theories has been introduced,  the corresponding field equations can be readily found. Taking variations with respect to the metric and the Ricci tensor, we obtain
\begin{align}\label{eq:variation}
\delta\mS
&=\frac{\lambdat^4}{\lambda^2}\int\d^4x\sqrt{-g}\left[
 \left(\frac{\partial F}{\partial M^{\mu}{}_\alpha}  g^{\mu\nu}+\frac{\partial F}{\partial N^{\mu}{}_\nu}g^{\mu\alpha}\right)\delta R_{\nu\alpha} \right.\nonumber\\
&\;\left.+ \left(\frac{\partial F}{\partial M^{\mu}{}_\alpha}  R_{\nu\alpha}+R_{\alpha\nu}\frac{\partial F}{\partial N^{\mu}{}_\alpha} -\frac{\lambda^2}{2}g_{\mu\nu}F\right)\delta g^{\mu\nu} \right].
\end{align}
Since the Ricci tensor is a function of the connection and does not depend on the metric, the metric field equations can be readily identified as 
\bea
\frac{\partial F}{\partial M^{(\mu}{}_\alpha}  R_{\nu)\alpha}+R_{\alpha(\nu}\frac{\partial F}{\partial N^{\mu)}{}_\alpha} -\frac{\lambda^2}{2}g_{\mu\nu}F=\frac{\lambda^2}{2\lambdat^4}T_{\mu\nu}
\label{metricFE1}
\eea
where we have added the energy-momentum tensor of the matter fields on the right hand side. We will assume that the matter fields are minimally coupled to gravity so that the energy-momentum tensor is defined as $T_{\mu\nu}\equiv -\frac{2}{\sqrt{-g}}\frac{\delta S_m}{\delta g^{\mu\nu}}$ where $S_m$ is the action of the matter fields. This energy-momentum tensor is conserved with respect to the Levi-Civita connection of the metric. This will be important because the matter field equations will decouple from the connection\footnote{This will be true at least for standard bosonic fields. Fermions, fields coupled to the curvature or Galileon-like fields couple directly to the connection and they would need to be treated separately. We are grateful to Claudia de Rham and Andrew Tolley for discussions on this.}. We can see that for the Einstein-Hilbert action with\footnote{Note that $[\Mm]=[\Nm]$.} $F=\frac{\lambda^2}{16\pi G\lambdat^4}[\Mm]$ we recover Einstein equations in the Palatini formalism $G_{(\mu\nu)}(\Gamma)=8\pi GT_{\mu\nu}$. Another class of theories worth a comment here are those for which the LHS of (\ref{metricFE1}) vanishes. This is the case for instance in Eddington's theory which has $F\propto\det(\Mm)$. Of course, this simply reflects the fact that the gravitational sector does not depend on the metric in this theory.

The metric field equations can be written in a more compact form by using matrix notation as
\be
\Big(\Fm_M \Rm^T+\Fm_N \Rm\Big)+\Big(\Fm_M \Rm^T+\Fm_N \Rm\Big)^T-\lambda^2\gm F=\frac{\lambda^2}{\lambdat^4}\Tm
\ee
where we have defined the matrices
\be
(F_M)_\mu{}^\nu\equiv\frac{\partial F}{\partial M^{\mu}{}_{\nu}}\quad \text{and} \quad(F_N)_\mu{}^\nu\equiv\frac{\partial F}{\partial N^{\mu}{}_{\nu}}.
\ee
Now we can use the definition of the matrices $\Mm$ and $\Nm$ in order to express the Ricci tensor in terms of them and, in addition, we can use the relation $\Nm=\gm^{-1}\Mm^T\gm$, so the above equation can alternatively be written as
\ba
\Fm_M\Mm^T+\gm\Mm\Fm_M^T\gm^{-1}+\Fm_N\gm\Mm\gm^{-1}+\Mm^T\gm\Fm_N^T\gm^{-1} \nonumber\\
= \lambda^2F+\frac{\lambda^2}{\lambdat^4}\Tm\gm^{-1}\,.
\label{eqM}
\ea
This (unusual) representation of the metric field equations puts forward that, at least in principle, one can use algebraic methods to solve for $\Mm$  in terms of the energy-momentum tensor and the metric, i.e., we can obtain $\Mm=\Mm(\Tm,\gm)$. As will be seen shortly, the formal dependence of  $\Mm$ on $\Tm$ and $\gm$ will be relevant in order to deal with the connection equation. 

It is interesting to note that the non-linearity of these equations will generically lead to the existence of several branches of solutions. In physical applications, however, one expects that only one branch is continuously connected with GR at low curvatures.  There is a particularly interesting case when we actually have $\Mm=\Mm(\Tm\gm^{-1})$, i.e., if we have a perfect fluid, $\Mm$ will be a function of the energy density and the pressure alone, with no explicit dependence on the metric components. Although this might seem quite stringent, it is actually the case for physically interesting situations, like cosmological and black hole scenarios. \\

Concerning the connection equations, we can use the expression for the variation of the Ricci tensor of an arbitrary connection given by
\be
 \delta R_{\nu\alpha}=\nabla_\lambda\delta\Gamma_{\alpha\nu}^\lambda -\nabla_\alpha \delta \Gamma_{\lambda\nu}^\lambda+2 \mathcal{T}_{\rho\alpha}^\lambda \delta \Gamma_{\lambda\nu}^\rho
\ee
with the torsion tensor
\begin{equation}
\mathcal{T}_{\rho\nu}^\lambda\equiv ( \Gamma_{\rho\nu}^\lambda-\Gamma_{\nu\rho}^\lambda)/2, 
\end{equation}
to finally obtain from (\ref{eq:variation}) the field equations, which become
\begin{eqnarray}
\nabla_\lambda\left( \sqrt{-g}W^{\nu\alpha}\right)-\delta_\lambda^\alpha \nabla_\rho\left( \sqrt{-g}W^{\nu\rho}\right)   & & \\
 - 2\sqrt{-g}\left(\mathcal{T}_{\kappa\lambda}^\kappa W^{\nu\alpha}-\delta^\alpha_\lambda \mathcal{T}_{\kappa\rho}^\kappa W^{\nu\rho}+\mathcal{T}_{\lambda\rho}^\alpha W^{\nu\rho}\right) &=&0 \nonumber
 \label{connectionEq}
\end{eqnarray}
where we have defined
\be
W_\mu{}^\alpha\equiv\Big(\Fm_M+\gm\Fm_N^T\gm^{-1} \Big)_\mu{}^\alpha
\label{defW}
\ee
and $W^{\nu\alpha}=g^{\nu\lambda}W_\lambda{}^\alpha$. Tracing over the indices $\lambda$ and $\alpha$, one finds an expression for $\nabla_\rho\left( \sqrt{-g}W^{\nu\rho}\right)$, which can be inserted back into (\ref{connectionEq}) to obtain
\begin{eqnarray} \label{eq:connec2}
\nabla_\lambda\left( \sqrt{-g}W^{\nu\alpha}\right)&=& 2\sqrt{-g}\left(\mathcal{T}_{\kappa\lambda}^\kappa W^{\nu\alpha}-\frac{1}{3}\delta^\alpha_\lambda \mathcal{T}_{\kappa\rho}^\kappa W^{\nu\rho}\right. \nonumber \\ &+&\left. \mathcal{T}_{\lambda\rho}^\alpha W^{\nu\rho}\right).
\end{eqnarray}
With further manipulations, one can show that the symmetric and antisymmetric parts of $W^{\nu\alpha}$ are coupled to each other through the object $\mathcal{T}_{\lambda\rho}^\nu+\frac{1}{3}\left(\delta^\nu_\rho \mathcal{T}_{\kappa\lambda}^\kappa-\delta^\nu_\lambda \mathcal{T}_{\kappa\rho}^\kappa\right)$. Setting that part of the torsion to zero, one gets two decoupled equations for $W^{(\nu\alpha)}$ and $W^{[\nu\alpha]}$ (see \cite{Olmo:2013lta} for a more detailed discussion on this point). One can then impose $W^{[\beta\nu]}=0$ (which implies that $R_{[\alpha\beta]}=0$) and regard its equation as a consistency relation for the torsion-free condition\footnote{Besides mathematical simplicity, setting the torsion to zero at the level of the field equations is physically well justified, though some comments are necessary in order to better understand its implications.  In this sense, a correspondence between the potential microscopic defects of the space-time structure and the affine degrees of freedom represented by the torsion and non-metricity has been recently established \cite{Lobo:2014nwa}. Following a well-known result from condensed matter systems, torsion can be associated with line-like defects in  crystals (dislocations and disclinations), while non-metricity accounts for point-like defects (vacancies and intersticials). This two types of defects have interactions between them and, therefore, both should be taken into account in a general discussion. However, there are physically relevant situations in which point-like defects dominate, being line-like defects negligible or simply absent. Setting the torsion to zero, therefore, is analogous to neglecting potential dislocations or disclinations in the microscopic space-time structure. This justifies why the field equations were derived assuming the {\it a priori} existence of torsion. Imposing vanishing torsion {\it a priori}, i.e., at the level of the action, would thus lead to a physically different theory in which line-like defects would not be allowed to exist. }. In that case, which is the one we are focusing in this work, the connection equation can be formally solved by  algebraic means, as will be shown shortly.

Under the above assumptions, therefore, the connection is determined by
\be\label{eq:connec1}
 \nabla_\lambda\left( \sqrt{-g}W^{\nu\alpha}\right)=0 \ ,
\ee
with $W^{\nu\alpha}$ symmetric. A key point to realize now is that, as pointed out following Eq.(\ref{eqM}), $\Wm$ can be written as an algebraic function of the matter content, through its energy-momentum tensor $\Tm$, and the metric tensor $\gm$. Hence, provided solutions to the algebraic equations can be found, the connection equations will also allow to solve  for the connection algebraically because the objects inside the covariant derivatives in (\ref{eq:connec1}) do not depend on the connection. Given that the connection can now be obtained by algebraic means rather than by solving differential equations, we conclude that the independent connection  in this type of Palatini theories does not introduce new dynamical degrees of freedom (under our assumptions). Therefore, there is no need to impose additional boundary conditions associated with extra modes.

The connection in Eq.(\ref{eq:connec1}) can be readily solved by introducing an auxiliary metric $\gt^{\mu\nu}$ satisfying
\be
\sqrt{\vert\gt\vert}\gt^{\mu\nu}=\sqrt{-g}W^{\beta\nu}.
\label{effectivemetric1}
\ee
This equation will hold for all the branches obtained as solutions of (\ref{eqM}) and we see that for each of the branches the signature of the auxiliary metric will coincide with that of the symmetric part of $W^{\mu\nu}$. There is no obvious reason why this matrix should have the same signature as the spacetime metric $g_{\mu\nu}$, although in the branch that matches GR at low densities (curvatures) $W^{\mu\nu}$ will have Lorentzian signature as it corresponds. If the signature changes at high curvatures, then $W^{\mu\nu}$ must become non-analytic at some point. 

By taking determinants in both sides of (\ref{effectivemetric1}), one finds that $\sqrt{\vert\gt\vert}=\sqrt{\vert g\vert}\sqrt{\vert\det\Wm\vert}$, which leads to
\be
\tilde{g}^{\alpha\beta}=\frac{1}{\sqrt{\vert \det \hat W\vert}}g^{\alpha\lambda}W_\lambda{}^\beta
\ee
or, equivalently
\be
 \tilde{g}_{\alpha\beta}=\sqrt{\vert \det \hat W\vert}\big(W^{-1}\big)_\alpha{}^\lambda \,g_{\lambda\beta} \ .
\label{effectivemetric2}
\ee
The Ricci tensor is thus given by the Levi-Civita connection of the auxiliary metric $\gt_{\mu\nu}$ and, as a consequence, it is symmetric. This further implies that for these solutions $\Mm$ and $\Nm$ coincide, which allows to write the field equations (\ref{eqM}) in a simpler form.
Furthermore, in many cases (and in particular for the case we will consider below) the symmetry of the Ricci tensor is inherited by $\Mm$ and, consequently, by $\Nm$. If we now remember the relation $\Mm=\gm^{-1} \Nm^T\gm$ we conclude that, for the solutions under consideration, we have $\Mm=\gm^{-1}\Mm\gm$ and, analogously $\Nm=\gm^{-1}\Nm\gm$, which means that they are invariant under a similarity transformation given by the metric tensor. This property is useful because then we have that\footnote{This follows from the fact that $\gm^{-1}\Fm_N\gm$ is a similarity transformation and, therefore, it is given by $\Fm_N$ evaluated at $\Mm=\gm^{-1}\Mm\gm$ and  $\Nm=\gm^{-1}\Nm\gm$. If we now use that $\Mm$ and $\Nm$ are invariant under a similarity transformation determined by $\gm$ for the considered solutions we finally obtain our result.} $\gm^{-1}\Fm_N\gm=\Fm_N$ and, therefore, we finally obtain from (\ref{defW}) the simpler expression $\Wm=\Fm_M+\Fm_N$. Equipped with this result we can then write the advertised simplified form of the metric field equations as
\bea
\Wm \Rm+\Rm\Wm^T=\lambda^2\left(F\gm +\frac{1}{\lambdat^4}\Tm\right).
\eea
Now, we can use (\ref{effectivemetric2}) to obtain $\Wm\Rm=\sqrt{\vert\det \Wm\vert} \gm\hat{\gt}^{-1}\Rm$ and, thus, express the above equation as
\be\label{eq:EOMGeral}
R^\alpha{}_{(\mu}g_{\nu)\alpha}=\frac{\lambda^2}{2\sqrt{\vert\det \Wm}\vert} \left(Fg_{\mu\nu}+\frac{1}{\lambdat^4}T_{\mu\nu}\right)
\ee
where we have defined $R^\alpha{}_\mu=\gt^{\alpha\lambda}R_{\lambda \mu}(\gt)$. When $\hat R$ and $\hat g$ are simultaneously diagonalizable, the product $R^\alpha{}_{(\mu}g_{\nu)\alpha}$ is symmetric and turns the metric field equations into
\be
R^\mu{}_\nu(\gt)=\frac{\lambda^2}{2\sqrt{\vert\det \Wm}\vert} \left(F\delta^\mu{}_\nu+\frac{1}{\lambdat^4}T^\mu{}_\nu\right)
\label{eqR}
\ee
with $T^\mu{}_\nu=g^{\mu\lambda}T_{\lambda\nu}$. This representation of the field equations has been previously obtained in different classes of theories in the Palatini formalism such as $f(R,R_{\mu\nu}R^{\mu\nu})$ theories \cite{Olmo:2009xy}, the Born-Infeld theory of gravity introduced in \cite{BF} (see \cite{Olmo:2013gqa} for details), $f(R)$ \cite{Makarenko:2014lxa,Makarenko:2014nca} and $f(|\hat\Omega|)$ \cite{Odintsov:2014yaa} deformations of the Born-Infeld theory and the extensions introduced in \cite{Jimenez:2014fla}.  Remarkably, in this representation the metric field equations resemble those of GR, but with a modified source term. Moreover, as commented before, for cases in which the matrix $\Mm$ can be expressed as a function of $\Tm\gm^{-1}$ (or, equivalently, $\gm^{-1}\Tm$), the RHS of this equation only depends on such a combination and, therefore, it can be regarded as a pure source term determined by the matter fields (i.e., the energy density and pressure in the case of a perfect fluid). This point was more explicitly discussed in \cite{Delsate:2012ky} for the case or pure Born-Infeld inspired gravity, where some of the consequences of these modified couplings to matter were explored. 

\section{Perfect fluid solutions}
After obtaining the general field equations, we will now consider a simple scenario where the matter source is given by a perfect fluid so that its energy-momentum tensor reads\footnote{Notice that our Ansatz is for $\Tm\gm^{-1}$, but since it is diagonal it would read the same for $\gm^{-1}\Tm$.}
\begin{equation}
{T_\mu}^{\nu}=\left(\begin{array}{cc} -\rho
 & \vec{0} \\
\vec{0} & p \Id_{3\times 3}
\end{array}\right)  \ .
\end{equation} 
Given the diagonal character of this ${T_\mu}^{\nu}$, we expect diagonal solutions for $\Mm$ and $\Nm$. Since we are assuming minimally coupled matter fluids, this stress-energy tensor satisfies the usual conservation equation (defined by the divergence of $T_{\mu\nu}$ with respect to the Levi-Civita connection of $g_{\mu\nu}$). In a spatially flat FLRW metric (see below), we thus get
\be
\dot{\rho}+3H(\rho+p)=0,
\ee
with $H$ the Hubble expansion rate. Of course, in addition to this equation we need some knowledge about the fluid equation of state to be able to obtain the evolution of the energy density.

From the relation $\Mm=\gm^{-1}\Nm^T\gm$ we deduce that $\Mm=\Nm$, which will be parametrized as 
\begin{equation}
{M^\mu}_{\nu}={N^\mu}_{\nu}=\left(\begin{array}{cc} M_0
 & \vec{0} \\
\vec{0} & M_1 \Id_{3\times 3}
\end{array}\right)  \ .
\end{equation} 
The metric field equations reduce to
\bea
\Big[(\Fm_M)_0{}^0+(\Fm_N)_0{}^0\Big]M_0=\frac{\lambda^2}{2}\left(F-\frac{\rho}{\lambdat^4}\right),\\
\Big[(\Fm_M)_i{}^i+(\Fm_N)_i{}^i\Big]M_1=\frac{\lambda^2}{2}\left(F+\frac{p}{\lambdat^4}\right),
\label{perfectfluideq}
\eea
where we are not summing over the repeated index $i$. As discussed above for the general case, these algebraic equations allow to solve for $M_0$ and $M_1$ in terms of $\rho$ and $p$. This allows to write $\Wm$ also in terms of $\rho$ and $p$ and solve the connection equation (\ref{eq:connec1}) introducing the auxiliary metric (\ref{effectivemetric2}).  

In order to illustrate how the space-time metric is obtained, let us consider a cosmological scenario where the line element is 
\be
\d s^2=-n(t)^2\d t^2+a(t)^2\d\vec{x}^2
\ee
with $n(t)$ and $a(t)$ the lapse and scale factor functions respectively. The lapse can be fixed by a time reparametrization, but we will keep it arbitrary for later convenience. Then, we will have that $\Wm=\text{ diag}(W_0,W_1,W_1,W_1)$ where $W_0$ and $W_1$ will be functions of $\rho$ and $p$ to be determined from the metric field equations (\ref{perfectfluideq}). Then, according to (\ref{eq:connec1}),  the auxiliary metric associated to the independent connection takes the form
\be
\d \tilde{s}^2=\sqrt{\vert W_0W_1^3\vert}\left(-\frac{n(t)^2}{W_0}\d t^2+\frac{a(t)^2}{W_1}\d\vec{x}^2\right) \ ,
\ee
which is again an FLRW metric with effective lapse and scale factor functions given by
\be
\nt(t)^2=\frac{n(t)^2\sqrt{\vert W_0W_1^3\vert}}{W_0}\quad\text{and}\quad\at(t)^2=a(t)^2\frac{\sqrt{\vert W_0W_1^3\vert}}{W_1}.
\label{effectiveFLRW}
\ee
We thus see that $W_0$ and $W_1$ must be positive in order to have the same signature for both metrics.

\section{Tensor perturbations}
We now compute the equations corresponding to tensor perturbations propagating on the background solution for a perfect fluid obtained in the precedent section. Since we are considering tensor perturbations, we will use the following Ansatz for the metric and energy-momentum tensor\footnote{We drop here the perfect fluid assumption and let the fluid has anisotropic stress that could source the metric perturbations.}
\begin{equation}
\delta g_{\mu\nu}=\left(\begin{array}{cc} 0
 & \vec{0} \\
\vec{0} & \delta g_{ij}
\end{array}\right)  \ , \hspace{0.5cm} \text{and} \hspace{0.5cm}  \delta T^\mu{}_\nu=\left(\begin{array}{cc} 0
 & \vec{0} \\
\vec{0} & \Pi^i{}_j
\end{array}\right) 
\end{equation} 
with $\delta g_{ij}\delta^{ij}=\Pi^i{}_i=0$. In this section, spatial indexes will be raised and lowered with the Kronecker delta $\delta_{ij}$. The conservation equation for the fluid does not allow to solve for the anisotropic stress at first order so it should be regarded as an external source (determined by the microscopic details of the fluid, in analogy to the equation state parameter for the background). We will assume that the structure of these perturbations will be inherited by the perturbations of all the remaining quantities. In particular, the perturbation in the matrix $\Mm$ must also be symmetric because we are considering purely tensor perturbations and an antisymmetric component will necessary come from an axial vector perturbation. 

A very important property shared by all the tensor perturbations that will greatly simplify our computations is the fact that any product of background matrices with one perturbation will be proportional to the perturbation. More explicitly, if $\bar{\hat{A}}$ and $\delta \hat{B}$ are a background and a perturbation matrices respectively, we have that $\bar{\hat{A}}\delta\hat{B}\propto\delta\hat{B}$. This is so because all the perturbations live on the hypersurfaces orthogonal to the time direction and the background quantities are proportional to the identity on such hypersurfaces. The importance of this remark is that all matrices commute at first order. Therefore, Eq. (\ref{eqM}) can be written, up to first order in tensor perturbations, as follows:
\be
\big(\Fm_M+\Fm_N\big)\Mm=\frac{\lambda^2}{2}\left(F\Id+\frac{1}{\lambdat^4}\Tm\gm^{-1}\right).
\label{eqdeltaM}
\ee
Thus, as we discussed above, the matrix $\Mm$ will be an algebraic function of the combination $\Tm\gm^{-1}$ up to first order for tensor perturbations, i.e., we will have that $\delta \Mm\propto\hat{\Pi}$ and, therefore, $\delta \Wm=\xi(\rho,p)\hat{\Pi}$  with $\xi(\rho,p)$ a zeroth order function of the background quantities that can be determined from Eq (\ref{eqdeltaM}), which, at first order, reads
\be
\delta\Wm\Mm+\Wm\delta\Mm=\frac{\lambda^2}{2\lambdat^4}\hat{\Pi}.
\label{eqdeltaM2}
\ee
An important feature to be noticed here is that if the fluid does not present anisotropic stresses, the perturbation $\delta\Wm$ vanishes at first order for tensor perturbations. This is the case for instance if the matter source is given by a perfect fluid.

Now, from the connection equation we again obtain that the connection is, to first order in tensor perturbations, the Levi-Civita connection of the auxiliary metric
\be
\tilde{g}_{\alpha\beta}= \sqrt{\vert\det \Wm\vert}\big(W^{-1}\big)_\alpha{}^\lambda \,g_{\lambda\beta} 
\ee
so that we can obtain the relation between the perturbations of both metrics and the anisotropic stress as follows:
\be
\delta \gt_{ij}=\frac{\sqrt{\vert\det \Wm\vert}}{W_1}\left(\delta g_{ij}-\frac{\xi(\rho,p)}{W_1}\Pi_{ij}\right).
\ee
This relation can be turned into a more apparent expression by introducing the more convenient metric perturbations $\delta g_{ij}=a^2h_{ij}$ and $\delta \gt_{ij}=\tilde{a}^2\hT_{ij}$, in terms of which the above equation reduces to
\be
\hT_{ij}=h_{ij}-\frac{\xi(\rho,p)}{W_1}\Pi_{ij}.
\label{htToh}
\ee
In particular, if we have no anisotropic stresses $\Pi_{ij}=0$, both metric perturbations coincide. This particular result and for the specific case of Born-Infeld inspired gravity was obtained in \cite{EscamillaRivera:2012vz}. As we have shown here though, the coincidence of both metric perturbations is a completely general feature for the class of theories described by the action (\ref{eq:action0}) and in the absence of torsion and anisotropic stresses. Furthermore, it can be easily understood from the fact that the induced metrics on the hypersurfaces orthogonal to the time direction (or equivalently $g_{ij}$ and $\gt_{ij}$) are conformally related up to first order in tensor perturbations.  For this, it is crucial that $\Pi=0$, which, as we showed above, implies that $\Wm$ does not acquire tensor perturbations. If $\Pi \neq 0$, then $\Wm$ will also have tensor perturbations which are responsible for the disformal term in  (\ref{htToh}).

Now that we have the relation between both metric perturbations, we can proceed to compute the evolution equation for tensor perturbations. For that, we will perturb Eq. (\ref{eqR}) to obtain the equation
\be
\delta R^\mu{}_\nu(\gt)= \frac{\lambda^2}{2\sqrt{\vert\det \Wm\vert}\lambdat^4}\Pi^\mu{}_\nu
\label{pertR}
\ee
where we have used again that tensor perturbations of scalar quantities vanish. This equation represents a remarkable result for tensor perturbations in the class of Palatini theories under consideration in this work. In order to appreciate it, it is convenient to remind here the corresponding equation for GR, which reads
\be
\delta R^\mu{}_\nu(g)= 8\pi G\,\Pi^\mu{}_\nu.
\ee
Thus, Eq. (\ref{pertR}) for the auxiliary metric tensor perturbation $\hT_{ij}$ is analogous to the usual GR equation for the spacetime metric $g_{\mu\nu}$ with an effective Newton's constant determined by the background fluid
\be
8\pi G_{\rm eff}\equiv \frac{\lambda^2}{2\sqrt{\vert\det \Wm(\rho,p)\vert}\lambdat^4}.
\label{Geff}
\ee
Equipped with the above results, we can very readily write the evolution equation for tensor perturbations as
\be
\ddot{\hT}_{ij}+\left(3\Ht(t)-\frac{\dot{\nt}(t)}{\nt(t)}\right)\dot{\hT}_{ij}-\frac{\nt(t)^2}{\at(t)^2}\nabla^2 \hT_{ij}=16\pi G_{\rm eff}\Pi_{ij},
\label{htequation}
\ee
in complete analogy with GR. This equation coincides with the more specific equation found in \cite{EscamillaRivera:2012vz} for the particular case of Born-Infeld inspired gravity. Again, here we have shown that this is in fact a completely general result for the class of theories described by (\ref{eq:action0}). Thus, in order to compute the evolution of tensor perturbations for any of such theories, only the background solutions are required. 

With the solution of (\ref{htequation}), one can proceed to obtain the solution for the space-time metric perturbations via (\ref{htToh}). At this point we can already identify the potential presence of instabilities. Firstly, we see that in order to avoid laplacian instabilities, we need the factor in front of the laplacian in the above equation to be positive. By looking at the background equations (\ref{effectiveFLRW}) we conclude that this is the case if $W_0$ and $W_1$ have the same sign. This is always the case for solutions in which $\gt_{\mu\nu}$ and $g_{\mu\nu}$ have the same signature. This condition also guarantees the hyperbolicity of the equation.  Furthermore, under this assumption, we can fix the time coordinate so that $\nt=\at$, i.e., we can use conformal time for the auxiliary metric $\gt_{\mu\nu}$. If we further introduce the rescaled perturbation $\tilde{\texttt{h}}_{ij}=\at \hT_{ij}$, Eq.(\ref{htequation}) reads
\be
\hTc''_{ij}-\left(\nabla^2 +\frac{\at''}{\at}\right)\hTc_{ij}=16\pi G_{\rm eff}\at\Pi_{ij},
\label{htequationconformal}
\ee
where $'$ denotes derivative with respect to the conformal time of $\gt_{\mu\nu}$. In this equation we clearly see the usual appearance of an effective (time-dependent) mass term. The difference with respect to GR lies in the fact that such a mass is determined here by the background evolution of the auxiliary metric $\gt_{\mu\nu}$ rather than by the space-time metric. Now one can translate all the usual results from GR to this class of theories. In particular, we see that it is the metric $\gt_{\mu\nu}$ which must be (quasi) de Sitter in order to generate a (nearly) scale invariant  spectrum from quantum fluctuations. Moreover, such a (nearly) scale invariant spectrum will be generated for the perturbation $\tilde{\texttt{h}}_{}ij$. For the same reason, it is possible to have a de Sitter background solution for the space-time metric and not generating a scale invariant spectrum of tensor perturbations \cite{BIinflation}. We need to note however, that we have assumed regularity in both metrics and that they have the same signature. For instance, it was discussed in  \cite{EscamillaRivera:2012vz} the presence of instabilities associated to points where the auxiliary metric $\gt_{\mu\nu}$ becomes singular, even if the evolution of $g_{\mu\nu}$ is perfectly regular.

%
%
%
%
%

\section{Conclusions}

In this work we have considered a general class of gravitational theories in the Palatini formalism which are described by an arbitrary function of the (inverse) metric tensor and the Ricci curvature. The relevant quantities can be represented by means of two fundamental matrices $\Mm$ and $\Nm$ which encode the two independent products between the inverse metric and the Ricci tensor. After defining and discussing the general properties of this class of theories we have derived the corresponding field equations for both the metric tensor and the connection. We have argued how the metric field equations can be regarded as an algebraic equation for $\Mm$ (or equivalently $\Nm$) in terms of the matter content and the spacetime metric tensor. Then, we have shown that, provided one can work out an explicit solution for $\Mm$, the connection field equations can be seen as a set of algebraic equations for the connection. Moreover, under the torsion-free assumption, the connection actually corresponds to the Levi-Civita connection of an auxiliary metric, whose associated field equations take the Einstein-like representation (\ref{eqR}) for an arbitrary Lagrangian of the form (\ref{eq:action0}).
As a simple application, we have studied solutions for perfect fluids and in cosmological scenarios. \\

For the cosmological background solutions we have derived the properties of tensor perturbations. An important result valid for our general class of theories is that, in the absence of anisotropic stresses in the matter sector, the tensor perturbations of both the space-time metric and the auxiliary metric that generates the connection are the same.  If the matter energy-momentum tensor presents anisotropic stresses, there is a disformal term between both metrics perturbations. We have also shown that the tensor perturbations for the auxiliary metric satisfy the same equations as one finds in GR, but with a modified and time-dependent Newton's constant and propagating on top of the background auxiliary metric. Then, we have been able to easily obtain conditions for the absence instabilities in the evolution of tensor perturbations. It would be interesting to extend our results for more general theories containing the full Riemann tensor or, equivalently, Lagrangians depending on the co-Ricci and homothetic tensors in addition to the Ricci tensor. However, new methods will be necessary since it will not be possible in general to obtain the connection by the algebraic procedure used in this work.

\acknowledgments
We would like to thank Claudia de Rham, Nima Khosravi, Tomi Koivisto and Andrew Tolley for fruitful discussions.
J.B.J. acknowledges the financial support of A*MIDEX project (n¡ ANR-11-IDEX-0001-02) funded by the ``Investissements dÕAvenir" French Government program, managed by the French National Research Agency (ANR), the Wallonia-Brussels Federation grant ARC No. 11/15-040 and  MINECO (Spain) projects FIS2011-23000 and Consolider-Ingenio MULTIDARK CSD2009-00064.
G.J.O. is supported by a Ramon y Cajal contract, the Spanish Grant No. FIS2011-29813-C02-02, the Consolider Program CPANPHY-1205388, and the Grant No. i-LINK0780 of the Spanish Research Council (CSIC). This work has also been supported by CNPq (Brazilian agency) through Project No. 301137/2014-5.


\end{document}